\documentclass[pre,twocolumn,showpcs,amsmath,amssymb]{revtex4}

\usepackage{graphicx}
\usepackage{dcolumn}
\usepackage{bm}

\begin{document}

\title{The bold/timorous walker on the trek from home}

\author{Maurizio Serva}

\affiliation{Dipartimento di Ingegneria e Scienze dell'Informazione e Matematica,
Universit\`a dell'Aquila, 67010 L'Aquila, Italy}
\affiliation{Departamento de Biof\'isica e Farmacologia, Universidade 
Federal do Rio Grande do Norte, 59072-970 Natal-RN, Brazil}

\date{\today}

\begin{abstract}
We study a one-dimensional random walk with memory.
The behavior of the walker is modified 
with respect to the simple symmetric random walk (SSRW) only when he is
at the maximum distance ever reached from his starting point (home).
In this case, having the choice to move farther or to
move closer, he decides with different probabilities.
If the probability of a forward step is higher then the probability of
a backward step, the walker is bold, otherwise he is timorous.
We investigate the asymptotic properties of this bold/timorous 
random walk (BTRW) showing that the scaling behavior
vary continuously from sub-diffusive (timorous) to super-diffusive (bold).
The scaling exponents are fully determined with a new mathematical
approach.
\end{abstract}

\pacs{05.40.-a, 02.50.Ey, 89.75.Da}
\maketitle


\section{Introduction}

In 1827, the botanist Robert Brown noticed that pollen grains in water perform
a peculiar erratic movement. Many decades later,  in 1905, Albert Einstein 
\cite{E} gave an explanation of the pollen {\it random walk} in terms of collisions
with the water moleculae relating the diffusion coefficient to observable quantities.
Indeed, Einstein was scooped by Louis Bachelier which five years before,
in his 1900 doctoral thesis \cite{B1} and in a following paper \cite{B},
arrived to similar conclusions.
Actually, Bachelier was interested in the motion of prices on French stock market, 
but (log-)prices move like pollen in water and their {\it random walk}
can be treated mathematically on the same ground.

This twofold origin of random walk as a probabilistic tool is illuminating, 
in fact, this utensil can be applied everywhere "a walker" 
(a particle, a cell, an individual, a price, a language,...) moves erratically
in such a way that its square displacement $x^2(t)$
increases in average according to $\langle x^2(t)\rangle \sim t$.

In its simpler version, the path of a random walk is the output of a succession 
of independent random steps. In this case, the scaling relation 
$\langle x^2(t) \rangle \sim t$
is immediate. Nevertheless, in most cases, this relation also holds if memory effect
in size and direction of the steps are present, the requirement is that memory
is short ranged and steps have not diverging length.

The scaling relation $\langle x^2(t)\rangle \sim t$ is traditionally associated to
the appellative {\it normal diffusion}, while {\it anomalous diffusion} corresponds to
a scaling $\langle x^2(t)\rangle \sim t^{2 \nu}$ with $\nu \neq 1/2$.
In particular sub-diffusive behavior corresponds to $\nu < 1/2$ and super-diffusive behavior
to $\nu > 1/2$.

There is a very large number of phenomena which exhibit anomalous diffusion
as well a variety of models which have been used to describe them, 
we refer to \cite{BG,BH,MK1,MK2,RKS} for a review of both.

Broadly, anomalous diffusion may arise via diverging steps length, 
as in L\'evy flights or via long-range memory effects as in
fractional Brownian motion and in self avoiding random walks. 
Diverging steps length and long-range memory 
are two different ways of violating the necessary conditions for the 
central limit theorem when applied to random walks.

Anomalous diffusion (super-diffusion) in L\'evy flights \cite{L} is the simple consequence 
of the fact that the length of the steps has a heavy-tailed probability distribution. 
This does not mean that the problem is trivial,
see for example \cite{AC} where the authors consider the interesting case in which diffusion 
is {\it strongly} anomalous ($\langle x^q(t)\rangle \sim t^{q \nu}$ with $\nu$ depending on $q$). 

Anomalous diffusion induced by long-range memory is the
non self-evident output of the self-interaction of the walker position
at different times;
the most celebrate example probably being the "true" self avoiding random 
walk introduced quite a long time ago \cite{APP} and later
rigorously studied (see \cite{T,TW} and references therein).
In this model, the exponent $\nu$ depends only on dimensionality.

In some case, the mechanism which gives origin to anomalous scaling can be different
for example special deterministic or random
environments (see for example \cite{CS,VS}) or multi-particle interactions \cite{LB}.

Since exact solution of non-trivial models with memory are quite difficult to obtain,
some effort has been made in this direction.
For example, in the elephant model \cite{ST}, the walker decides the direction 
of his step depending on his previous decisions.
Unfortunately, this model, given the direction of the first step,
can be exactly mapped in a Markovian model, without necessity 
of enlarging the phase space, and, more importantly, the anomalous scaling is not
a consequence of an anomalous diffusion but of the movement of the center of
mass of the probability distribution of the position.
Some generalization of this model has been proposed (see \cite{BF}) which are genuinely
non-Markovian but which show the same problem concerning the
origin of the anomalous scaling.

In an other analytically treatable model it is considered the case of
a semi-Markovian sub-diffusive processes  in which the waiting time for a step
is given by a probability distribution with a diverging mean value \cite{SK}.

Random walks with memory have been also employed to model
the spreading of an infection in a medium with a history-dependent susceptibility
\cite{DB,DAB}, the focus, in this case, is the time scaling of the survival
probability (a trap is collocated somewhere) and not the scaling of diffusion.
Moreover, random walks with memory have been used in finance as, for example,
in \cite{BP}. In this paper it is described the strategy of a prudent investor 
which tries to maximize the invested capital while never decreasing his standard of life.
In \cite{DB,DAB} and in \cite{BP}, as in the model presented in this paper, the behavior of the walker
is modified only when it is at the maximum distance from the origin 
and Markovianity is recovered only when the phase space is properly enlarged.

Motivated by the scarcity of exact solutions,
we present in this paper a model which is treatable, one-dimensional, 
genuinely non-Markovian and which shows anomalous scaling
ranging from sub-diffusion to super diffusion according to a single continuous parameter.

The paper is organized as follows: in Section II we present the model
and we expose our results; in Section III we describe the
decomposition of the dynamics which is at the basis of our mathematical approach;
the asymptotic behavior is computed in Section IV; in section V
we write and exactly numerically solve the associated forward Kolmogorov equation;
Section VI contains our conclusions.
Some of the calculations whose result is used in Section III are postponed in a final Appendix.

\section{Model and results}
The model presented in this paper is one-dimensional, steps all have the 
same unitary length, time is discrete and the walker can only move left
or right at any time step.
The behavior of the random walker is modified 
with respect to the simple symmetric random walk (SSRW) only when he is
at the maximum distance ever reached from his starting point (home).
In this case, he decides with different probabilities
to make a step forward (going farther from home) or a step backward
(going closer to home). 
 
More precisely, the model is the following: 
the walker starts from home ($x(0)=0$), then, 
at any time he can make a (unitary length) step to the right or to the left 
\begin{equation}
x(t+1)=x(t)+\sigma(t)
\label{x}
\end{equation}
with $\sigma(t)= \pm 1$. We define 
\begin{equation}
y(t)= \max_{0 \le s \le t} |x(s)|
\label{y}
\end{equation}
which is the maximum distance from home he ever attained which obviously 
implies $-y(t) \! \le \! x(t) \! \le \! y(t)$. Then we assume  
\begin{itemize}
\item
$\sigma(0)= \pm 1$ with equal probability, i.e. the walker 
choses with equal probability the direction of the first step,
\item
$\sigma(t)= \pm 1$ with equal probability if the walker is not
at is maximum distance from home, i.e., $|x(t)|<y(t)$, 
\item
$\sigma(t)= {\rm sign}(x(t))$ with probability $p(y(t))$ and 
$\sigma(t)= -{\rm sign}(x(t))$ with probability $1-p(y(t))$ 
if $|x(t)|=y(t)$,
\item
the probability $p(y)$ depends on $y$ according to 
$p(y)= y^\gamma/(1+y^\gamma)$.
\end{itemize}
Therefore, simple symmetric random walk (SSRW) holds when $|x(t)|<y(t)$
but when the walker is at the maximum distance from home ($|x(t)|=y(t)$),
he boldly prefers to move farther if $\gamma > 0$ or timorously prefers to 
move closer if $\gamma < 0$.

Our goal is to find the asymptotic behavior of $y(t)$ and $|x(t)|$.
We preliminarily observe that in case $\gamma=0$ one has $p(y)=1/2$ 
which implies SSRW holds everywhere, also if the walker is at maximum distance.
In this case, ordinary scaling applies:
$\langle y^\alpha (t) \rangle \sim \langle|x(t)|^\alpha \rangle
\sim t^{\alpha/2}$ for any real positive $\alpha$ (the sign $\sim$ 
indicates that the ratio of the two sides asymptotically tends to a 
strictly positive constant. We use the sign $\simeq$ for
the stronger statement that the ratio tends to 1).

Results of this paper cam be summarized as follows:
\begin{itemize}
\item
$\langle y^\alpha (t) \rangle \simeq \langle y(t) \rangle^\alpha \simeq 
(t/2\nu)^{\alpha \nu } $ with  $\nu=1/(2-\gamma)$ for $ -\infty < \gamma < 0$,
\item
$\langle y^\alpha (t) \rangle \sim t^{\alpha \nu}$ with $\nu=1/(2-2\gamma)$
for $0 \le \gamma \le 1/2$,
\item
$\langle y^\alpha (t) \rangle \simeq \langle y(t) \rangle^\alpha \simeq t^\alpha$ 
for $1/2 < \gamma < \infty$.
\end{itemize}
Moreover, $\langle  |x(t)|^\alpha \rangle \sim \langle  y^\alpha (t)\rangle$ 
for all $\gamma$.

In both regions $ -\infty < \gamma < 0$ and $1/2 < \gamma < \infty$ 
relations are indicated with $\simeq$, i.e. 
the ratio of the two sides tends to 1 in the limit $t \to \infty$
providing both the scaling exponent and the scaling factor for
$\langle y^\alpha (t) \rangle$.
Furthermore, $\langle y^\alpha (t) \rangle \simeq \langle y(t) \rangle^\alpha$ 
which implies that the  variable $y(t)$ scales deterministically as its average. 

In particular, in region  $ -\infty < \gamma < 0$ the behaviour is sub-diffusive, 
as a consequence of the propensity the walker has to step in the home direction
when at maximum distance,
while, in the region $1/2 < \gamma < \infty$, behavior is ballistic,
with coefficient 1, as a consequence of the strong propensity to step away from
home when at maximum distance. 
\begin{figure}[!ht]
\includegraphics[width=3.truein,height=2.truein,angle=0]{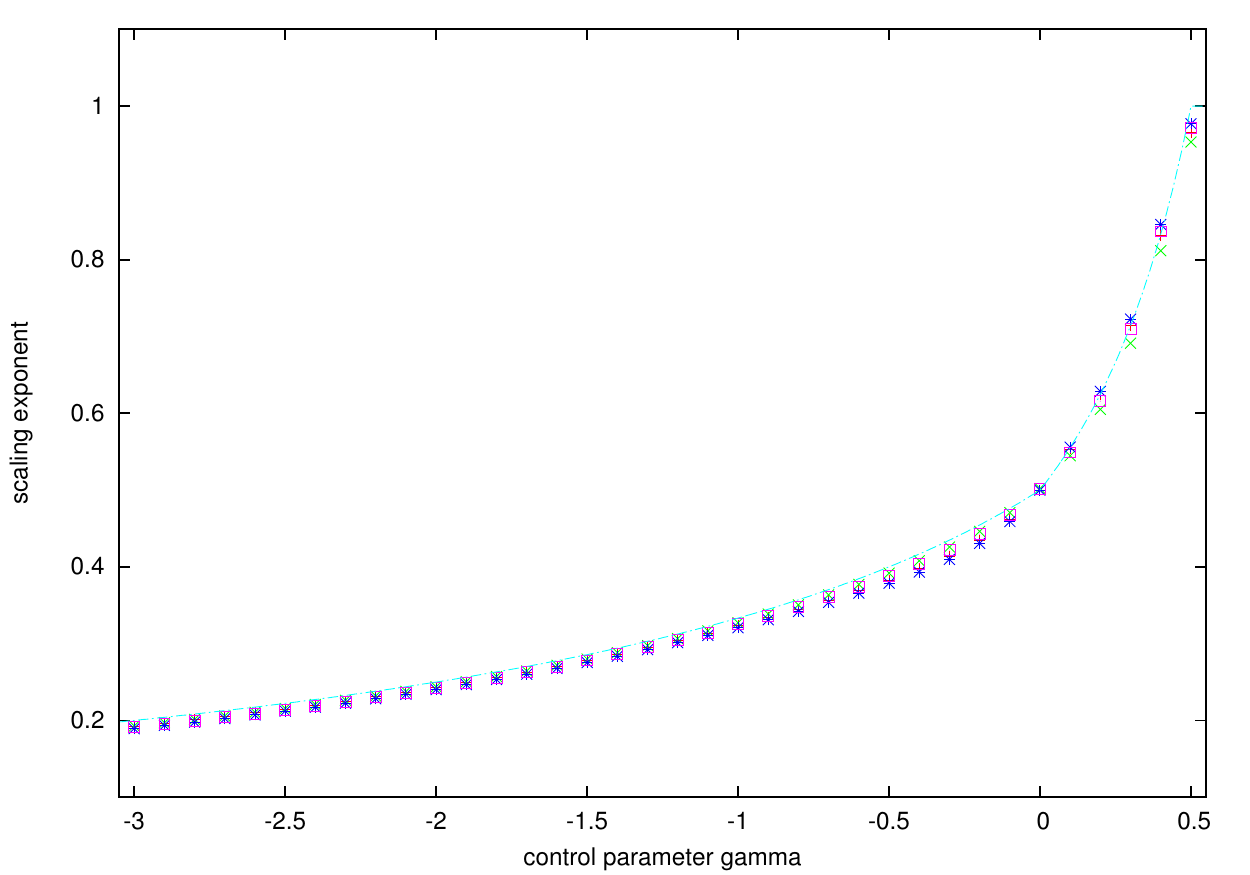}
\caption{Scaling exponent $\nu$ deduced from  $\langle |x(t)| \rangle$
(crosses, red), $\langle y(t) \rangle$ (slanted crosses, green),  
$\langle |x(t)|^2 \rangle$ (stars, blue)
and  $\langle y^2(t) \rangle$ (squares, violet) against 
prevision (full line).}
\label{fig1}
\end{figure}

Finally, in the intermediate region $0 \le \gamma \le 1/2$ only
the scaling exponent of $\langle y^\alpha (t) \rangle$ is determined.
If $\gamma = 1/2$, ordinary diffusion holds (with standard coefficients),
while in the region $0 < \gamma \le 1/2$, we have non-ballistic
super-diffusive behaviour as a consequence of the (not too strong)
propensity the walker has to step away from home when at maximum distance. 

The behavior of the anomalous scaling exponent $\nu$ with respect to the control parameter 
$\gamma$ in region $\gamma < 1/2$ is depicted in Fig. 1. 
In region $\gamma \ge 1/2$ the exponent $\nu$ equals 1, i.e. behavior is ballistic
(notice that, by construction, it cannot be super-ballistic).

Next three sections are devoted to the validation of the results here presented.

\section{Decomposition of the dynamics}
Here we outline the the decomposition of the dynamics which
is at the basis of our new mathematical approach.

Trajectories are decomposed in active journeys and lazy journeys. 
The lazy journey starts at the time $t$ when the walker leaves the maximum 
and it ends when he reaches it again at time $t+m+1$, i.e. $|x(t)|=y(t)=y$,
$|x(t+s)|< y$ for $1 \le s \le m$ and $|x(t+m+1)|=y$. 
The total number of steps of this journey is $1+m$ since the first step is for leaving 
the maximum and $m$ is the random number of steps necessary to reach it again
starting from a position $|x|=y-1$.
During all steps of the lazy journey the maximum remains the same.
The minimum duration of the lazy journey is two time steps 
($1+m=2$ when the walker immediately steps back to the maximum after having left it).

The active journey starts at the time $t+m+1$ when the walker 
arrives on a maximum and it ends when he leaves it at time $t+m+n+1$, 
i.e. $|x(t+m)|=y-1$, $|x(t+m+1+s)|= y+s$ for $0 \le s \le n$ and 
$|x(t+m+n+2)|=y+n-1$ (the first step of a new lazy journey). 
The total (random) number of time steps of this journey is $n$ 
with a minimum duration of 
zero steps ($n=0$ when the walker immediately leaves the maximum after being arrived).
During the active journey the maximum increases from $y$ to $y+n$.

A cycle journey is composed by a lazy journey followed by an active journey,
its duration is $1+m+n$ and the maximum increases of $n$. 

Notice that both $n=n(y)$ and $m=m(y)$ are random variables whose distribution only 
depends on $y$. In fact $m(y)$ is the SSRW first hitting time of one of the barriers
$y$ or $-y$ starting from position $x=y-1$ or $x=-y+1$, while 
the statistics of $n(y)$ is determined by $y$ through $p(y)$. 

We start by evaluating the probability $\pi (n|y)$ that the walker makes 
at least $n$ steps during the active journey,  i.e.
$\pi (n|y) = {\rm prob}\,(n(y) \ge n)$. Straightforwardly:
\begin{equation}
\pi (n|y)= \prod_{s=0}^{n-1} p(y+s) 
\label{pi}
\end{equation}
where $ p(y+s)= (y+s)^\gamma/(1+(y+s)^\gamma)$.

For large $y$, we have to distinguish three different ranges of $\gamma$:
\begin{itemize}
\item[$i)$]
$-\infty <\gamma < 0$, in this case $\pi (n|y) \le y^{n\gamma}$ where
the approximate equality $\pi (n|y) \simeq y^{n\gamma}$
holds for $n$ small with respect to $y$,
\item[$ii)$]
$0 <\gamma < 1$,  in this case $\pi (n= \beta y^\gamma|y) 
\simeq e^{-\beta}$, which means that $n(y) \simeq \xi y^\gamma$
where $\xi$ is a random variable distributed according
to an unitary exponential probability,
\item[$iii)$]
$ \gamma > 1$, in this case $\pi(n|y) \simeq e^{-\psi(y,n)}$
where $\psi(y,n) = \sum_{s=0}^{n-1} 1/(y+s)^\gamma$, noticeably, $\pi (\infty|y)$ 
is finite which implies that $n(y)$ is infinite with finite probability. 
\end{itemize}
The above relations are derived in the Appendix.

In both cases $i)$ and $ii)$ the average is 
$\langle n(y) \rangle \simeq y^\gamma$, while in case $iii)$ it diverges,
moreover, the standard deviation in case $i)$ is $\sigma_{n(y)} \simeq y^{\gamma/2}$
while in case $ii)$ is $\sigma_{n(y)} \simeq y^{\gamma}$.

The statistical properties of $m(y)$ are well known \cite{B,F}, in fact, 
as already stressed, 
$m(y)$ is simply the SSRW time for hitting one of the frontiers of the interval 
$[-y,y]$ starting from position $y-1$ (or $-y+1$).
Using standard martingale approach, it is easy to compute
the average $\langle m(y) \rangle \simeq 2y$ and the standard deviation
$\sigma_{m(y)} \simeq (8/3)^{1/2} y^{3/2}$ where the approximations hold
for large $y$ (see the Appendix).

We underline that standard deviation is larger than average and that
both diverge when $y \to \infty$, nevertheless, $m(y)$ has a finite probability
to be of order 1 (consider that $m(y) =1$ with 
probability $1/2$ independently on $y$).
This is reflected in the fact that the averages $\langle m(y)^{\beta} \rangle$
are of order 1 when $\beta$ is negative. For $m$ not very small, 
$\theta(m|y) \sim m^{-1/2}$ \cite{B,F} 
with a cutoff at $m_c \sim y^2$ where $\theta(m|y)$ drops to 0
($m_c$  is the typical time the walker can "see" the far barrier).
This implies that $\theta(m|y)$ is approximately a truncated L\'evy distribution. 
 
\section{Asymptotic analysis}

We have seen in previous section that in a cycle 
journey starting from a maximum $y$, time increases of is $1 + m(y) + n(y)$ and the
maximum increases of $n(y)$.

Let us indicate with $k$ (to be not confused with time $t$)
the progressive number identifying cycle journeys, 
each composed by a lazy journey followed by an active journey.
Also, let us indicate with $y(k)$ the value of the maximum when the cycle journey 
number $k$ starts.

The time $t$ is linked to the progressive number $k$ by the stochastic relation
\begin{equation}
t(k+1)= t(k)+ 1+ m(y(k))+n(y(k))
\label{tk}
\end{equation}
while the value of the maximum by
\begin{equation}
y(k+1)= y(k)+n(y(k))
\label{yk}
\end{equation}
where $m(y(k))$ and $n(y(k)$ are all independent random variables whose 
statistical properties we have already described.

In principle one should simply solve the two equations and, by substitution,
obtain the scaling behavior of $y(t)$. Obviously, this asks for some work.

We start our asymptotic analysis by considering the region $ -\infty <\gamma < 1$.
Let us consider first equation (\ref{yk}).
In the region of $\gamma$ we are considering, the variables
 have average $\langle n(y(k)) \rangle \simeq y(k)^\gamma$, 
then, from equation (\ref{yk}), one has
$\langle y(k+1)^{1-\gamma} \rangle \simeq \langle y(k)^{1-\gamma} \rangle
+(1-\gamma)$.
The omitted terms are of lower order in $y(k)$ since the standard deviation 
of the $n(y(k))$ can be $\sigma_{n(y(k))} \simeq y(k)^{\gamma/2}$
(for $-\infty < \gamma <0$) or $\sigma_{n(y(k))} \simeq y(k)^{\gamma}$ 
(for $0< \gamma <1$).
By integration we obtain $\langle y(k)^{1-\gamma} \rangle \simeq (1-\gamma) k$
and by iteration $\langle y(k)^{l(1-\gamma)} \rangle \simeq (1-\gamma)^l k^l$ 
where $l$ is a positive integer number.
Finally, by analytical continuation we have
$\langle y(k)^{\alpha} \rangle \simeq (1-\gamma)^{\alpha/(1-\gamma)}
k^{\alpha/(1-\gamma)} \simeq \langle y(k) \rangle^{\alpha}$ for any real $\alpha$. 
We have thus proven the relation 
\begin{equation}
y(k) \simeq (1-\gamma)^{1/(1-\gamma)} \, k^{1/(1-\gamma)} 
\label{ykf}
\end{equation}
which holds deterministically, i.e. fluctuations are comparatively negligible 
in the large $y(k)$ limit.

Let us now consider equation (\ref{tk}), by simple sum we get
\begin{equation}
t(k)= y(k)+ k+ M((k))
\label{tkf}
\end{equation}
where $y(k)$ is given by (\ref{ykf}) and $M(k)= \sum_{i=0}^{k-1} m(y(i))$
which is a sum of independent variables distributed according
to truncated L\'evy distributions with $\langle m(y(i)) \rangle \simeq 2y(i)$ 
and  $\sigma_{m(y(i))} \simeq (8/3)^{1/2} y(i)^{3/2}$.
According to (\ref{ykf}) we obtain the average 
\begin{equation}  
\langle M(k) \rangle \simeq \frac{2}{2-\gamma} \,
[(1-\gamma)  k]^{(2-\gamma)/(1-\gamma)}
\label{mk}
\end{equation}
and the standard deviation 
$\sigma_{M(k)}\sim  k^{(2-\gamma/2)/(1-\gamma)}$.

\smallskip
If $\gamma <0$, the standard deviation $\sigma_{M(k)}$ is asymptotically
negligible with respect to the average  $\langle M(k) \rangle$
and we can thus replace  $M(k)$ with its average $\langle M(k) \rangle$
in (\ref{tkf}). Furthermore we can neglect the smaller terms 
$y(k)$ and $k$ and obtain $t(k) \simeq \langle M(k) \rangle$ 
which solved with respect to $k$ and
substituted in (\ref{ykf}) finally gives the relation
\begin{equation}
y(t) \simeq (1-\gamma/2)^{1/(2-\gamma)} \, t^{1/(2-\gamma)}  
\label{yt}
\end{equation}
which holds deterministically in the region $-\infty < \gamma <0$.

On the contrary, if $0<\gamma < 1$ the standard deviation of $M(k)$ is larger 
than its average. In this case, it is necessary to determine the behavior of 
its probability distribution.
This can be done considering that all the independent $m(y((i))$ 
in the sum which defines $M(k)$ are distributed according to a truncated L\'evy.
Then, according to the generalized central limit for leptokurtic variables
\cite{L},
\begin{equation}
L(k)= \frac{1}{k^2} \sum_{i=1}^k m(y(i))
=\frac{1}{k^2} M(k)
\label{lk}
\end{equation}
is also a truncated leptokurtic variable, notice, in fact, 
that the denominator equals the power two of the number of the 
summed variables (L\'evy is $\alpha =1/2$ stable).
Also notice that $L(k)$ has average $ \sim k^{\gamma/(1-\gamma)}$
and variance $\sim k^{(3/2) \gamma/(1-\gamma)}$ which both diverge
in the large $k$ limit (truncation disappears).
Accordingly, $L(k)$ is of order 1 with probability 1 and all the 
averages $\langle L(k)^\beta \rangle$ with negative $\beta$ are of order 1.
This property is true for any $k$ and it also holds in the limit
$k \to \infty $ where $L(k) \to L$. 

Then, in the region $0 < \gamma < 1$, equation (\ref{tkf}) can be 
rewritten as
\begin{equation}
t(k) \simeq y(k) +k^2 L
\label{tl}
\end{equation}
where the term $k$ has been be dropped
since it is smaller both with respect to $y(k)$ and $k^2 L$. 

The region $0 < \gamma < 1$ splits into two subregions,
when $1/2 < \gamma < 1$, the term $y(k)$ is larger than $k^2 L$, therefore
we can assume $t(k) \simeq  y(k)$ and therefore $y(t) \simeq t$
holds deterministically.

When $0 < \gamma < 1/2$, on the contrary, $k^2 L$ is larger than $y(k)$
so that $t(k) \simeq k^2 L$. This implies $k=(t(k)/L)^{1/2}$ which
substituted in (\ref{ykf}) finally gives the relation
\begin{equation}
\langle y^\alpha (t) \rangle \simeq  \langle L^{-\alpha \nu} \rangle
 \, (1-\gamma)^{\alpha/(1-\gamma)} \, t^{\alpha \nu}
\label{ytfl}
\end{equation}
with $\nu= 1/(2-2\gamma)$. The average 
$\langle L^{-\alpha \nu} \rangle$ is of order 1
since the exponent is negative, but we are unable to determine 
its exact value in terms of $\gamma$ and $\alpha$. So we simply conclude 
that $\langle y^\alpha (t) \rangle \sim t^{\alpha \nu} $.

At this point only remains the region $1 < \gamma <\infty$,
but for these values of $\gamma$,
active walks of infinite length have a finite probability, 
therefore, after some excursions away from the maximum the walker decides 
once of all to follow the same direction remaining always on the maximum.
Accordingly, the relation $y(t) \simeq |x(t)| \simeq t$ holds deterministically.
Our analysis of the scaling behavior of $y(t)$ is thus concluded.

Since the walker spends part of the time on the maximum
where $|x(t)| =y(t)$ and part of the time in ordinary random walk 
in the interval $-y(t) < x(t) < y(t)$ where in average $|x(t)| \ge y(t)/2$
(the lazy trek always starts form the frontier), we also conclude 
that $\langle  |x(t)|^\alpha \rangle \sim \langle  y^\alpha (t) \rangle$.

\begin{figure}[!ht]
\includegraphics[width=3.truein,height=2.truein,angle=0]{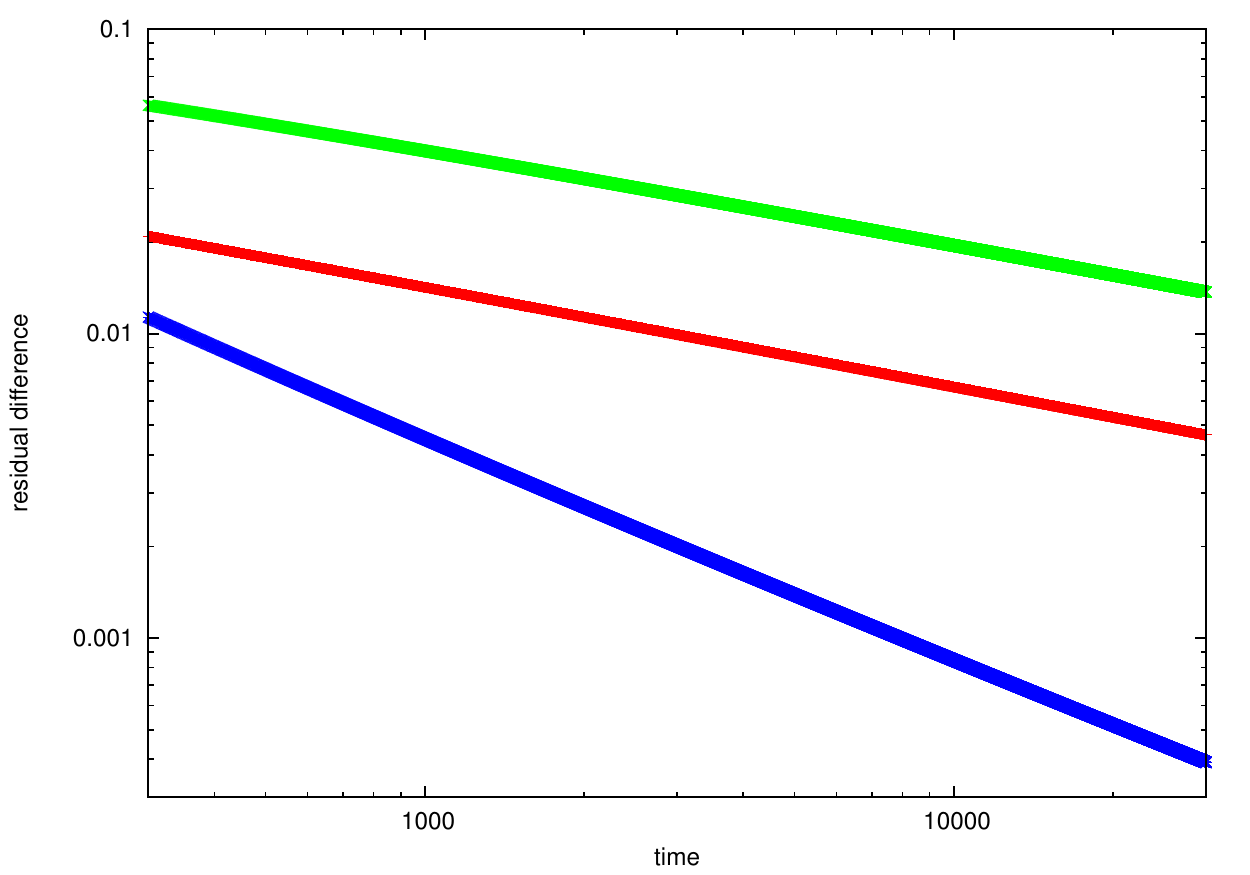}
\caption{Log-log plot of 
$\sqrt{\langle y^2(t) \rangle}/\langle y(t) \rangle -1$ 
(intermediate line, red),
$ \langle y(t) \rangle/  (t/2\nu)^{\nu} -1$ 
(upper line, green),
$2[\langle y(t+1) \rangle -\langle y(t) \rangle]/(t/2\nu)^{\nu-1} -1$
(lower line, blue) for $\gamma=-1$.}
\label{fig2}
\end{figure}

\section{Forward Kolmogorov equation}
We would like to test our results against the output
of the exact numerical solution of the forward Kolmogorov equation.  
Indeed, the process is non-Markovian, but it can be rendered Markovian by 
enlarging the phase space including the variable $y(t)$, and, therefore,
jointly considering the evolution of the variables $x(t)$ and $y(t)$.
Accordingly, it is possible to write a forward Kolmogorov equation for
$P(x,y,t)$ which is the probability that $x(t)=x$ and $y(t)=y$.

The initial condition ($t=0$) is $P(0,0,0)=1$ while all others $P(x,y,0)$ 
equal zero. 
First of all, notice that $P(x,y,t)$ obviously vanishes when $|x|>y$ 
and when $y>t$. Moreover, the symmetry of both initial condition and dynamics 
implies $P(x,y,t)=P(-x,y,t)$ for all $x$, $y$ and $t$.
Thus, let us write the forward equation only for $x \ge 0$.

At any time $t \ge 1$ one has $P(0,0,t)=0$, furthermore one has
$P(1,1,t)=0$ at even times, $P(1,1,t)=(1/2)^{(t+1)/2}$ at odd times, 
$P(0,1,t)=0$ at odd times and $P(0,1,t)=(1/2)^{t/2}$ at even times.

Assuming that $y \ge 2$, the forward Kolmogorov equation is completed by
\begin{equation}
P(x,y,t\!+\!1)=\frac{1}{2}P(x\!+\!1,y,t)+\frac{1}{2}P(x\!-\!1,y,t)
\label{k1}
\end{equation}
which holds when $0 \le x \le y -2$. In case $x=0$, 
we can use the symmetry to replace $P(-\!1,y,t)$ with $P(1,y,t)$
in the right hand side of the equation. When $x=y-1$ we have
\begin{equation}
P(y-1,y,t\!+\!1)=(1\!-p(y)) \,P(y,y,t)+\frac{1}{2}P(y\!-\!2,y,t)
\label{k2}
\end{equation}
where $p(y)= y^\gamma/(1+y^\gamma)$ and, finally, when $x=y$ we have
\begin{equation}
P(y,y,t\!+\!1)=p(y\!-\!1)P(y\!-\!1,y\!-\!1,t)+\frac{1}{2}P(y\!-\!1,y,t).
\label{k3}
\end{equation} 

We have numerically exactly solved the Kolmogorov equation
and computed $\langle y(t) \rangle$, $\langle |x(t)| \rangle$,
$\langle y^2(t) \rangle$ and $\langle |x(t)|^2 \rangle$.
The scaling exponent can be obtained by the ratio
$ \log(\langle y(t) \rangle)/\log(t)$ and analogous expressions where
$\langle y(t) \rangle$ is replaced by $\langle |x(t)| \rangle$,
$\langle y^2(t) \rangle$ or$\langle |x(t)|^2 \rangle$.
Nevertheless, convergence is much faster 
if one computes the exponent from 
$ \log(\langle y(t_2) \rangle/\langle y(t_1) \rangle)/\log(t_2/t_1)$
and analogous expressions since the scaling factor is wiped out.
In Fig. 1 we plot our results for $t_2= 11000$ 
and $t_1= 10000$ against prevision.
Independently of the use of $\langle y(t) \rangle$, $\langle |x(t)| \rangle$,
$\langle y^2(t) \rangle$ or $\langle |x(t)|^2 \rangle$ 
we find excellent agreement.
Notice that for $\gamma=1/2$ the exponent $\nu$ equals 1 (ballistic behavior),
for larger values of $\gamma$ necessarily it must have the same value,
thus, without loss of information, Fig. 1 ends at $\gamma=1/2$. 

In the region $-\infty < \gamma < 0$
we also have found the explicit scaling factor and 
the relation $\langle y^\alpha (t) \rangle \simeq \langle y(t) \rangle^\alpha$.
In order to confirm the latter we consider
the residual difference $\sqrt{\langle y^2(t) \rangle}/\langle y(t) \rangle -1$
up to 30000 time steps.
This difference converges to 0 according to a power law as shown 
by the log-log plot in Fig. 2 for two orders of magnitude of time.
Moreover, the log-log plot of
$ \langle y(t) \rangle/  (t/2\nu)^{\nu} -1$
shows power law convergence to 0 (Fig. 2) proving that both 
scaling factor and scaling exponent are correct.
A faster power law convergence (Fig. 2) can be obtained considering
$2(\langle y(t+1) \rangle -\langle y(t) \rangle)/(t/2\nu)^{\nu-1} -1$,
since only the differential average at the largest time contributes.
Plot in Fig. 2, corresponds to the case $\gamma=-1$ but we have verified 
the same power law behavior for various values in the region $-\infty < \gamma < 0$.

\section{Conclusions}
Anomalous diffusion in this model is induced by long-range memory
in a conceptually very simple manner,
furthermore, the model is one-dimensional and it is controlled 
by a single parameter.
In spite of this conceptual simplicity, the scaling behavior 
unfold all possibilities varying continuously from sub-diffusive to ballistic.
More precisely, if the walker timorously prefers to go back
when it is at the frontier of unexplored regions, it is sub-diffusive, 
on the contrary, if he boldly prefers to go where he never has gone
before, it is super-diffusive. 

The sub-diffusive region is below the threshold $\nu =0$. 
Above the threshold $\nu =1/2$ the process is
ballistic, and the walker moves uniformly at constant velocity.
Finally, in the region above the threshold $\nu =0$ but below the threshold $\nu =1/2$
the process is super-diffusive but sub-ballistic.
This region is probably the most interesting
since the walker has an intermittent behavior,
with bursts of linear growth, followed by longer bursts of
random motion. This behavior is typical of the transition from laminar 
to turbulent behavior in chaotic systems \cite{PV}. 

This reach phenomenology can be used, in principle, to model a variety of phenomena.
We think, for example, to the problem of foraging strategies, with the 
walker (animal) changing his attitude when he is at the frontier of unexplored regions. 
The aim, in this case, is to evaluate the degree of success of the search
in comparison with ordinary random walk search and L\'evy search \cite{VA}.
Also in epidemics, recent focus is on the effects of super-diffusive spreading
of an infection, via heavy-tailed distributed jumps \cite{SMA}.
The present model could be alternative, with super-diffusive (or sub-diffusive) 
spreading arising as an effect of memory of infection agents.
Moreover, the orthography of languages performs a random walk 
on the discrete space of possible vocabularies \cite{SP,PS}. As in present model,
the jump rates are different if changes are in the direction
of a radical innovation or if they run on an already treaded territory.
Finally, as we already mentioned, the (non-ballistic) super-diffusive region of the
present model could represent a stochastic counterpart to
chaotic systems with intermittent behaviour (see also \cite{CM}).

We conclude pointing out that the
mathematical characterization of the BTRW in this paper is far to be complete, for example,
all the scaling factors for the variable $x(t)$ and the scaling factors 
for the variable $y(t)$ in the region $0 < \gamma \le 1/2$
remain unknown as well all the correlations at different times among variables.

Finally, we would like to underline
that this model could be successfully extended to higher dimensions.

\section*{ACKNOWLEDGEMENTS}
The author warmly thanks Michele Pasquini which greatly contributed
both to the  ideation of the model and its analytical and numerical
solution. Definitely, this research would have not been accomplished
without his contribution. 

The present work was partially supported by PRIN 2009 protocollo 
n. 2009TA2595.02.

\section*{APPENDIX}
In the first part of this Appendix
we prove the three relations i), ii) and iii) of Section III. 

In the region $-\infty <\gamma < 0$, one has 
$p(y+s)= (y+s)^\gamma/(1+(y+s)^\gamma) \le y^\gamma$
which implies $\pi (n|y) \le y^{n\gamma}$.
If $n$ is small with respect to $y$, 
one also has $ p(y+s)= (y+s)^\gamma/(1+(y+s)^\gamma) \simeq y^\gamma$
which, using (\ref{pi}), immediately gives 
the approximated equality $\pi (n|y) \simeq y^{n\gamma}$.

In the region $0 <\gamma < 1$, we directly obtain from (\ref{pi}), 
\begin{equation}
[p(y)]^{n} \le \pi (n|y)= \prod_{s=0}^{n-1} p(y+s)
\le  [p(y+n)]^{n},
\label{app1}
\end{equation}
in fact, being $\nu$ positive,
$p(y)$ is the smallest among the elements of the product
and $p(y+\beta y^\gamma)$ the largest.

Then assume $n=\beta y^\gamma$  (if $\beta y^\gamma$ is not
an integer then $n=\beta y^\gamma+\epsilon$ where $ 0 <\epsilon <1 $),
one immediately gets
\begin{equation}
[p(y)]^{\beta y^\gamma+\epsilon} \le \pi (n= \beta y^\gamma|y)
\le  [p(y+\beta y^\gamma)]^{\beta y^\gamma+\epsilon}
\label{app2}
\end{equation}

Then, using the definition of $p(y)$ and taking into account that
$0 <\gamma < 1$, it is straightforward
to verify that  the limit for $y \to \infty$ of both bounds
is $e^{-\beta}$ so that
\begin{equation}
\pi (n= \beta y^\gamma|y) \simeq e^{-\beta}
\label{app3}
\end{equation}
The above approximated equality means that $n(y) \simeq \xi y^\gamma$
where $\xi$ is a random variable distributed according
to an unitary exponential probability.

Finally consider the region $ \gamma > 1$, we have
\begin{equation}
p(y+s)= 1/(1+(y+s)^{-\gamma}) \simeq e^{-1/(y+s)^\gamma}
\label{app4}
\end{equation}
which implies $ \pi(n|y) \simeq e^{-\psi(y,n)}$
where $\psi(y,n) = \sum_{s=0}^{n-1} 1/(y+s)^\gamma$.
Noticeably, $\nu > 1$ implies that $\pi (\infty|y)$ is finite which ,
in turn, implies that $n(y)$ is infinite with finite probability. 
 
We compute now the average and standard deviation of
$m(y)$ which appear in Section III.

We preliminary remark that the process $x(t)$
is a SSRW when it is not on the maximum, therefore
$m(y)$ is simply the random time necessary for hitting one of the 
frontiers of the interval $[-y, y]$
starting from position $y-1$ (or $-y+1$). 

Assume that at time $t+1$ the walker is in $y-1$,
i.e. $x(t+1)=y-1$ (the choice $x(t+1)=-y+1$ is symmetrical 
and leads to the same results).
Also assume that the walker hits for the first time one of the barriers at 
time $t+1+m(y)$, i.e. $x(t+1+m(y))=y$ or $x(t+1+m(y))=-y$.
Strong martingale property implies that the average of $x(t+1+s) -x(t+1)$
equals zero at any non-negative $s$ also if $s$ is a random time, therefore
\begin{equation}
\langle x(t+1+m(y)) \rangle -x(t+1) = 0
\label{app5}
\end{equation}
Given that $a$ is the probability of hitting the barrier
$y$ and $1-a$  is the probability of hitting the barrier $-y$,
one has that $\langle x(t+1+m(y)) \rangle = a y+(1-a)(-y)$.
Therefore, equality (\ref{app5}) rewrites $a y-y(1-a)- y+1 =0$ which,
in turn,  implies $a=1-1/(2y)$.

Then we notice that $[x(t+1+s) -x(t+1)]^2-s$ is also a martingale, therefore
\begin{equation}
\langle [x(t+1+m(y)) -x(t+1)]^2 - m(y) \rangle = 0 
\label{bpp2}
\end{equation}
where $\langle [x(t+1+m(y)]^2 \rangle = a y^2+(1-a)y^2=y^2$
which implies $\langle m(y) \rangle = a+(1-a)(2y-1)^2 \simeq 2y$.

Moreover, $[x(t+1+s) -x(t+1)]^4-3s^2$ is as well a martingale and, therefore, 
\begin{equation}
\langle [x(t+1+m(y)) -x(t+1)]^4 - 3 m^2(y) \rangle = 0 
\label{bpp3}
\end{equation}
where  $\langle [x(t+1+m(y)]^2 \rangle = a y^4+(1-a)y^4=y^4$
which implies $\langle m^2(y) \rangle = (a+(1-a)(2y-1)^4)/3 \simeq (8/3)y^3$.

The standard deviation can be finally easily computed as
$\sigma_{m(y)} =[\langle m^2(y) \rangle - 
\langle m(y) \rangle^2]^{1/2}\simeq (8/3)^{1/2} y^{3/2}$.

\end{document}